\documentclass[sigconf]{acmart}

\AtBeginDocument{%
  }

\setcopyright{none}
\acmDOI{10.48550/arXiv.2411.17391}
\acmISBN{}

\acmConference[1st International Workshop on Low Carbon Computing]{1st International Workshop on Low Carbon Computing}{December 3,
  2024}{Glasgow, Scotland}

\usepackage{todonotes}
\usepackage{soul}
\usepackage[super]{nth}

\begin{document}


\title[The belief in Moore's Law is undermining ICT climate action]{The belief in Moore's Law is undermining ICT climate action}

\author{Adrian Friday, Christina Bremer, Oliver Bates and Christian Remy}
\affiliation{%
  \institution{Lancaster University}
  \city{Lancaster}
  \country{United Kingdom}
}

\author{Srinjoy Mitra}
\affiliation{%
  \institution{University of Edinburgh}
  \city{Edinburgh}
  \country{United Kingdom}
}
\email{srinjoy.mitra@ed.ac.uk}

\author{Jan Tobias M\"uhlberg}
\affiliation{%
  \institution{Universit\'e Libre de Bruxelles}
  \city{Brussels}
  \country{Belgium}
}
\email{jan.tobias.muehlberg@ulb.be}

\renewcommand{\shortauthors}{Friday et al.}



%


\begin{abstract}
The growth of semiconductor technology is unprecedented, with profound transformational consequences for society. This includes feeding an over-reliance on digital solutions to 
systemic problems such as climate change (`techno-solutionism').  Such technologies come at a cost: environmental, social and material.  We unpack topics arising from ``The True Cost of ICT: From Materiality to Techno-Solutionism (TCICT)'', a workshop held at the International ICT for Sustainability (ICT4S) conference 2024 in Stockholm, Sweden---exploring
, as a matter of global climate injustice
, the drivers and material dependencies of these technologies. We point to the importance of addressing ICT's impacts as a system, rather than purely in terms of efficiency and energy use. 
We conclude by calling 
to build a community of like-minded and critical colleagues 
to address the intersectional climate impacts of the semiconductor industry and the techno-solutionism it embodies.
\end{abstract}

\maketitle

\section{Introduction}

In the last 50 years, semiconductor technology has unquestionably enjoyed unprecedented growth compared to any other industrial sector, from 2000 components per semiconductor chip in the 1970s to over 50 billion today~\cite{mcgregor-true22}. This trend that was already observed by Gordon Moore in 1975, who stipulated a bi-annual doubling of transistors in integrated circuits, which has manifested massive gains in computational power and efficiency; and simultaneously underwritten revolutions in digital mediated industries such as communication, transportation, 
and 
latterly, of course, sponsoring 
the rebirth of artificial intelligence and particularly deep learning.

Digital industrialisation over the last 50 years has touched most aspects of business and society, leading for some to a quasi-religious faith that technology can address many key societal challenges we face today, including but not limited to, climate change.  Such solutions bringing about an apparent `technological utopia' in which social and environmental challenges are solved through better technology, models, digital twins, and the decarbonisation and dematerialisation 
of other industries.

In contrast, we hypothesise that the techno-solutionist paradigm---the never-ending cycle of innovation in digital/semiconductor technologies 
---is 
dangerous
~\cite{johnston2020techno, saetra2023technology}. 
Neglecting the globally significant and growing material, carbon and social footprints of ICT 
\textit{in the present} while dreaming of solutions \textit{for the future}.

In this position paper we argue the case for the growing impacts of ICT, where the past 50 years have shown no guarantee that long term energy and material consumption will ever go down, despite massive gains in efficiency---\textit{a classic rebound effect}---more powerful and efficient ICT ultimately results in a net gain in devices with even larger total energy and material consumption~\cite{coroamua2019digital,lange2020digitalization}. 
We argue for a more nuanced and responsible view of the benefits and costs of ICT in climate solutions, especially in the Global North.

\section{What are the true costs of ICT?}

Techno-solutionism is the belief that there are technological solutions to all problems faced by humanity, even where the problem has originated from our over-reliance on technology itself. This is a narrative that is particularly prevalent, though not exclusively found in the Global North, but that deeply permeates society.  
Mainly through technology, it is argued, we could achieve a sustainable utopia, full of economic growth and affluence, that does not cause undue harm~\cite{jones2023, mills2021cloud}. There is a widespread belief among businesses, policymakers and the general public, that it is mainly through technological innovation that climate change can be solved.  Relentless ICT innovation (epitomised by Moore’s Law) is probably a key driver behind this ideology~\cite{mitra2023roleictinnovationperpetuating}. 
We argue this optimism is unfounded and \textit{actively impedes} more decisive, meaningful and immediate action on climate (or societal) change.

To explore the breadth of ICT's impacts and the concept and drivers of techno-solutionism more deeply, we held the ``The True Cost of ICT: From Materiality to Techno-Solutionism'' workshop\footnote{\url{https://ict4s24-tcict.github.io/}} at the International ICT for Sustainability (ICT4S) conference, on Monday the 
\nth{24} of June 2024, in Stockholm, Sweden. Attendees were required to submit short position statements, from which the chairs invited short talks on assessing ICT's impacts, paired with guest speakers on specific topics of social and global justice, and studies of communities' relationship with mining and mineral resource extraction. The hybrid-format workshop attracted over 30 researchers from both academia and industry with an interest in ICT sustainability, at different career stages and with a wide range and depth of experience.  Field notes were taken by the authors during discussions, with breakout discussions captured on paper and online using physical and virtual post-it notes.  The lead authors synthesised these using a simple bottom up thematic analysis.  While the workshop talks and breakout sessions covered far more than we can represent here, we zoom in on specific aspects of ICT's impacts drawn from the resulting workshop discussions that are sometimes missed in one-dimensional accounts focusing on energy or greenhouse gas (GHG) emissions.


\subsection{The best known costs of ICT}

In 2021, estimates placed the externality costs of ICT in terms of GHG as being equivalent to global air travel~\cite{freitag2021real}. 
Although, there is considerable controversy not least shrouded in a mysterious game of non-disclosure of metrics relating to growth and resource consumption by major digital infrastructure providers in the absence of significant government policy. 

One underlying narrative is that data centres are no cause for concern as they are achieving ever higher efficiency rates~\cite{masanet.e-re-2020}.  Another, that each hardware generation brings increases in performance per unit energy~\cite{malmodin2024ict}. While, these are undoubtedly true---as data centres increase in scale, so efficiencies relating to amortising running costs increase; and similarly as transistor densities grow (in line with Moore's Law~\cite{schaller1997moore}), so we can argue that overall energy budgets due to CPUs/GPUs and cooling should fall.  However, this increase in capability also feeds economic and market growth for new ICT products and infrastructures, leading to further higher capacity including networks and data centres.  
Large AI companies accelerating data centre growth have even overshot their self-imposed emission targets~\cite{milmo.d-insatiableai-2024}. 

According to the International Energy Agency, data centres, cryptocurrencies, and AI consumed about 460 TWh of electricity worldwide in 2022, almost 2\% of total global electricity demand; they also predict that global electricity demand from data centres could double towards 2026~\cite{iea2024}. This puts specific pressure on electricity grids: with Microsoft, Amazon and others' facilities in Ireland forecast to consume a third of the country’s energy by 2026 and already 53\% of the country's renewable energy supply~\cite{bloomberg-ai-havoc-2024}.

\subsection{The lesser known costs of ICT}

Centering the narrative on efficiency gain, plays nicely with existing market 
drivers towards more capability, and 
more product sales. Nevertheless, it decentres  
less talked about \textit{material costs} of ICT. 
The production of ICT equipment consumes materials, and the faster digital technology becomes embedded in other products and services, the more material consumption and reliance on material extractivist practices underpins this.

ICT has perhaps uniquely complex supply chains, depending on sometimes vary rare minerals that exist globally in tiny quantities~\cite{fitzpatrick2015conflict}. This raises particular pressures in parts of the world where these materials are found. Geo-political challenges with this have also driven a recent focus on sovereignty of production and resilience~\cite{chips-america-2024}. 
The mounting challenge of ever higher transistor counts and increasing throughput of chips, places growing reliance on even less abundant parts of the periodic table~\cite{sun2020summarizingcpugpudesign}.


Large scale computing facilities, such as hyperscalar data centres, are now sufficiently large energy consumers that they place major burdens on energy grids and drive major energy projects through power purchase agreements~\cite{moss-3mile-2024}.  This can reduce energy resilience and increase the cost of energy for communities~\cite{ortar2022powering}; but it also can displace other energy users who can't afford to compete for this capacity~\cite{bloomberg-ai-havoc-2024}. 
It is important to recognise that creating renewable energy infrastructures is also not free from energy and material dependencies, especially globally!

\subsection{Human, social cost and new injustices}


ICT has indirect links to extractivist practices such as mining and waste handling, some with questionable labour practices and consequences to human health and for environmental degradation~\cite{saha2021electronic}.  A significant failure of the technology industry is the relatively low rates of recycling (as low as 20\%), helping drive this. 

Water use is emerging as an important datacentre concern; new metrics like `water use effectiveness' (WUE) aim to address this, but like PUE, talk of a race to improve a specific ratio rather than reduce absolute consumption.  
This could be said to ignore headline issues like the overall rate of growth, and environmental sensitivity where this impact occurs.  Using water where it is abundant is clearly less of a concern than using it where it is already scarce and takes away from populations who rely upon it~\cite{mollen.a-go-2024}.

What of populations displaced from lands where these precious minerals lie, such as the indigenous S\'{a}mi in the Nordics~\cite{mollen.a-go-2024}?  Who has the power and the money to compete with global tech giants? And what of the damage to the peoples and biology due to the use of chemicals and machinery to reach them~\cite{grant2013health}?  The continued injustice 
from the rapid growth and adoption of ICT based solutions in the Global North, on populations in the Global South~\cite{mitra2023need,mollen.a-go-2024} reprises neocolonialism.  If ICT energy demand 
looks set to consume `unreasonable proportions' of renewable energy supply~\cite{gupta-chasing-2021}, already outstripping anticipated demand in net zero roadmaps---shouldn't this cause us to ask what is `a reasonable share' to dedicate to ICT in our future?

\subsection*{Call to action}

For too long the ICT sector has been complacent or negligent to its global and environmental impacts.  This feeds the narrative that efficiency gains and replacing old with new (more efficient) technology (as characterised by Gordon Moore's famous observation on the transistor doubling rate) is sufficient to address the massive and growing climate and environmental burdens of global ICT. Whereas, efficiency gain without limit is in effect an engine of growth.  We need to urgently address impacts of green-tech `solutions' avidly promoted by technology companies (and supported by governments in the Global North) as a matter of climate injustice~\cite{sanderson2022volt, mitra2023need}---moving the debate from the three `E's of energy, emissions and efficiency, to one that centres international justice and the full scale of societal and environmental harms.

We call to form an inclusive community of like-minded and critical researchers and practitioners to work to challenge current conceptions of how `progress' in this industry helps to fuel this reliance on the promise of future green technologies (e.g., renewable energy, electric vehicles, carbon capture, geoengineering); whilst ignoring immediate consequences of technology to climate change, downplaying less techno-centric nature-based solutions. 
How might we engage industry stakeholders and beyond to shift to more benign and massively longer lasting technologies that respect exploited nature and peoples?  Closer to home, paradoxically, can we recognise as educators and mentors how deeply technology education is steeped in techno-solutionism and ideas of `innovation at any cost', producing generations of technologists equally trained not to question the environmental and social costs of what they produce~\cite{mitra-paradoxinhe-2023}?

\begin{acks}
We thank our funder, EPSRC (Grant ref.\ EP/T025964/1), and all the wonderful workshop participants at ICT4S in Stockholm. We gratefully acknowledge the Brussels-Capital Region---Innoviris for financial support under grant numbers 2024-RPF-2 and 2024-RPF-4.
\end{acks}

\balance

\bibliographystyle{ACM-Reference-Format}
\bibliography{sample-base}

\end{document}